\documentclass[reprint,
 amsmath,amssymb,
 aps,
floatfix,
showkeys
]{revtex4-1}

\usepackage[USenglish]{babel}
\usepackage{slashed}
\usepackage{physics}
\usepackage{graphicx}
\usepackage{dcolumn}
\usepackage{bm}
\usepackage[usenames]{color}
\usepackage{xcolor}
\usepackage{float}

\newcommand{\be}{\begin{equation}}
\newcommand{\ee}{\end{equation}}
\newcommand{\bea}{\begin{eqnarray}}
\newcommand{\eea}{\end{eqnarray}}
\newcommand{\beas}{\begin{eqnarray*}}
\newcommand{\eeas}{\end{eqnarray*}}

\newcommand{\ps}{\slashed{p}}

\begin{document}

\title{Magnetic field dependence of the neutral pion mass in the linear sigma model with quarks: The strong field case}


\author{Alejandro Ayala$^{1,2}$, Jos\'e Luis Hern\'andez$^{1}$, L. A. Hern\'andez$^{1,2,3,4}$, Ricardo L. S. Farias$^5$, R. Zamora$^{6,7}$ }
\affiliation{%
$^1$Instituto de Ciencias Nucleares, Universidad Nacional Aut\'onoma de M\'exico, Apartado Postal 70-543, CdMx 04510, Mexico.\\
$^2$Centre for Theoretical and Mathematical Physics, and Department of Physics, University of Cape Town, Rondebosch 7700, South Africa.\\
$^3$Departamento de F\'isica, Universidad Aut\'onoma Metropolitana-Iztapalapa, Av. San Rafael Atlixco 186, C.P, CdMx 09340, Mexico.\\
$^4$Facultad de Ciencias de la Educaci\'on, Universidad Aut\'onoma de Tlaxcala, Tlaxcala, 90000, Mexico.\\
$^5$Departamento de F\'isica, Universidade Federal de Santa Maria, Santa Maria, RS 97105-900, Brazil.\\
$^6$Instituto de Ciencias B\'asicas, Universidad Diego Portales, Casilla 298-V, Santiago, Chile.\\
$^7$Centro de Investigaci\'on y Desarrollo en Ciencias Aeroespaciales (CIDCA), Fuerza A\'erea de Chile, Casilla 8020744, Santiago, Chile.}%


\begin{abstract}

We use the linear sigma model with quarks to find the magnetic field-induced modifications to the neutral pion mass at one-loop level. The magnetic field effects are introduced by using charged particle propagators in the presence of a magnetic background in the strong field regime. We show that when accounting for the effects of the magnetic field on the model couplings, the vacuum sigma field and the neutral pion self-energy, the neutral pion mass decreases monotonically as a function of the field strength.   We find an excellent qualitative and quantitative agreement with recent lattice QCD calculations, reproducing the monotonically decreasing trend with the field strength as well as the decrease when lattice data approaches the physical vacuum pion mass from larger values.

\end{abstract}


\maketitle

\section{Introduction}\label{sec1}
Electromagnetic fields play a relevant role for the dynamics of strongly interacting systems. For instance, it is well known that an external magnetic field helps to catalyze the breaking of chiral symmetry, producing a stronger quark-antiquark condensate~\cite{catalysis}. On the other hand, when temperature is taken into account, magnetic fields inhibit the condensate formation producing the opposite effect whereby the pseudocritical temperature for the chiral phase transition is reduced, it is the so-called inverse magnetic catalysis~\cite{LQCD,Bruckmann,Farias,Ferreira,Ayala0, Ayala1,Ayala2,Ayala3,Avancini,Ayala4,vertex1,vertex2,qcdcoupling,Mueller,imcreview}. In this context, the properties of hadron degrees of freedom in the presence of magnetic fields have become a subject of intense study~\cite{bali01,iranianos,simonov03,aguirre02,tetsuya,dudal04,kevin,gubler,noronha01,morita,morita02,sarkar03,band,Ayalachi,nosso1,nosso03,zhuang,iran,scoccola01,huang01,scoccola02,scoccola03,luch,farias01,mao01,sarkar,sarkar02,zhang01,huan02,simonov01,fraga01,aguirre,taya,shinya,andersen01,kojo,simonov02,ghoshrho,AMMmesons,TBspectralprop,ghoshEPJA,Avila,Avila:2020ved,dudal,dudal02,andrei02,he,nucleon,barionslattice,Sheng}.

Given that, from the hadron sector, the dynamics of chiral symmetry breaking is dominated by pions, it becomes important to study the influence of magnetic fields on pion properties such as masses and form factors. On general grounds, charged and neutral pions behave differently under the influence of an external magnetic field. Charged pions with mass $m_0$ and at rest in the direction of the magnetic field, have an energy spectrum given by $E^2=m_0^2+(2n +1)|eB|$, where $|eB|$ is the field strength and $n$ labels the $n$-th Landau level. The lowest energy state can be interpreted as the magnetic field-dependent mass, which is then given by $m_B^2=m_0^2+|eB|$. In contrast, neutral pions do not experience directly the effects of a magnetic background and thus their mass remains at first sight unaffected. Interactions with other particles that populate the strongly interacting vacuum can change this picture. In fact, the magnetic field driven modifications of the neutral pion mass were first computed by lattice QCD (LQCD) calculations in Refs.~\cite{Hidaka,Luschevskaya}. These works found contradictory results: whereas Ref.~\cite{Luschevskaya} obtains a neutral pion mass that monotonically decreases with the field strength, Ref.~\cite{Hidaka} finds a dip at an intermediate value and then an increase for larger field strengths. This discrepancy was analyzed in Refs.~\cite{Endrodi,Ding} where the monotonically decrease of the pion mass as a function of the field strength was confirmed.

The problem has been also addressed from the point of view of effective models. Working within the linear sigma model with quarks (LSMq) in the weak field limit, Ref.~\cite{Ayala1} has shown that the neutral pion mass starts off decreasing as a function of the field strength. An important element for this finding is to account for the magnetic field driven modification of the boson self-coupling. Within the same model, and also with magnetic field modified couplings, Ref.~\cite{Das} finds that the neutral pion mass starts off decreasing to then increase at an intermediate value of the field strength, becoming even larger than the mass at zero field strength. A similar behavior is found in Refs.~\cite{Li,Scoccola} using the NJL model. 

The results of Ref.~\cite{Das}, which are in contrast with the most recent LQCD results~\cite{Endrodi,Ding}, may be due to the use of a procedure, advocated in Ref.~\cite{Scoccola}, to remove the vacuum, which in Ref.~\cite{Ayala1} has been shown to not describe the large magnetic field strength limit, generating a discrepancy with the well established LLL result. Moreover, since the effective boson-fermion coupling in Ref.~\cite{Das} is computed from the magnetic field corrections to the quark mass, it is not clear what, if any, is the role of Schwinger's phase factor when this effective coupling is computed from the one-loop {\it triangle} perturbative correction.

In order to assess whether the use of one-loop magnetic field-modified couplings can account for the behavior of the recent LQCD results for the magnetic field dependence of the neutral pion mass, in Ref.~\cite{AHHFZ} we made a detailed study of the magnetic field modifications to the boson self-coupling and boson-fermion coupling in the LSMq. We found that the couplings experience a monotonic decrease as a function of the field strength. The pending question is thus to clarify what are the overall ingredients that can explain the neutral pion mass monotonic decrease as a function of the field strength. In this work we address this question within the LSMq, showing that the elements driving the neutral pion mass behavior are the properly combined effects of the magnetic field corrections to the couplings together with the contribution from charged particles to the one-loop pion self-energy and to the vacuum expectation value of the sigma field.

The work is organized as follows: In Sec.~\ref{sec2}, we introduce the linear sigma model with quarks. In  Sec.~\ref{sec3} we make a quick survey of the way magnetic field effects are introduced into the propagators of charged bosons and fermions. In  Sec.~\ref{sec4}, we compute the necessary elements to obtain the magnetic modification of the pion mass to one-loop order, namely, the neutral pion self-energy, the vacuum expectation value of the sigma field from the effective potential, and the correction to the couplings. In Sec.~\ref{sec5} we compute the magnetic corrections to the neutral pion mass and compare to recent LQCD calculations, showing that the monotonic decrease with the field strength can be reproduced. We finally summarize and conclude in Sec~\ref{sec6}. We reserve for the appendices the explicit calculation details for the one-loop corrections to both, the neutral pion self-energy and the effective potential.

\section{Linear Sigma Model with quarks} \label{sec2}
The LSMq is an effective model that describes the low-energy regime of QCD, incorporating the spontaneous breaking of chiral symmetry. The Lagrangian for the LSMq can be written as 
\begin{eqnarray}
\mathcal{L}&=&\frac{1}{2}(\partial_{\mu}\sigma)^{2}+\frac{1}{2}(\partial_{\mu}\vec{\pi})^{2}+\frac{a^{2}}{2}(\sigma^{2}+\vec{\pi}^{2})-\frac{\lambda}{4}(\sigma^{2}
+\vec{\pi}^{2})^{2}\nonumber\\
&+&i\bar{\psi}\gamma^{\mu}\partial_{\mu}\psi-ig\gamma^{5}\bar{\psi} \vec{\tau} \cdot \vec{\pi}\psi-g\bar{\psi}\psi\sigma.
\label{lagrangian}
\end{eqnarray}
Pions are described by an isospin triplet,  $\vec{\pi}=(\pi_1,\pi_2,\pi_3)$. Two species of quarks are represented by an $SU(2)$ isospin doublet, $\psi$. The $\sigma$ scalar is included by means of an isospin singlet. Also, $\lambda$ is the boson self-coupling and $g$ is the fermion-boson coupling. $a^2>0$ is the mass parameter.

To allow for spontaneous symmetry breaking, we let the $\sigma$ field develop a vacuum expectation value $v$
\begin{eqnarray}
    \mathcal{L}&=&\frac{1}{2}\partial_{\mu}\sigma \partial^{\mu}\sigma+\frac{1}{2}\partial_{\mu}\pi_{0}\partial^{\mu}\pi_{0}+\partial_{\mu}\pi_{-}\partial^{\mu}\pi_{+}\nonumber\\
    &-&\frac{1}{2}m_{\sigma}^{2}\sigma^{2}-\frac{1}{2}m_{0}^{2}\pi_{0}^{2}-m_{0}^{2}\pi_{-}\pi_{+}+i\bar{\psi}\slashed{\partial}\psi\nonumber\\
    &-&m_{f}\bar{\psi}\psi+\frac{a^2}{2}v^2-\frac{\lambda}{4}v^4+\mathcal{L}_{int},
    \label{linearsigmamodelSSB}
\end{eqnarray}
where the charged pion fields can be expressed as
\begin{equation}
 \pi_\pm=\frac{1}{\sqrt{2}}(\pi_1\pm i\pi_2),
\end{equation}
and the interaction Lagrangian is defined as
\begin{equation}
\begin{split}
    \mathcal{L}_{int}&=-\frac{\lambda}{4}\sigma^{4}-\lambda v\sigma^{3}-\lambda v^{3}\sigma-\lambda\sigma^{2}\pi_{-}\pi_{+} -2\lambda v \sigma\pi_{-}\pi_{+}\\
    &-\frac{\lambda}{2}\sigma^{2}\pi_{0}^{2}-\lambda v\sigma \pi_{0}^{2}-\lambda \pi_{-}^{2}\pi_{+}^{2}-\lambda\pi_{-}\pi_{+}\pi_{0}^{2}-\frac{\lambda}{4}\pi_{0}^{4}\\ 
    &+a^{2}v\sigma -g\bar{\psi}\psi\sigma-ig\gamma^{5}\bar{\psi}\left(\tau_{+}\pi_{+}+\tau_{-}\pi_{-}+\tau_{3}\pi_{0}\right)\psi.
    \label{interactinglagrangian}
\end{split}    
\end{equation}
In order to include a finite vacuum pion mass, $m_{0}$, one adds an explicit symmetry breaking term in the Lagrangian of Eq.~\eqref{linearsigmamodelSSB} such that
\begin{equation}
    \mathcal{L}\rightarrow \mathcal{L'}=\mathcal{L}+\frac{m_{0}^{2}}{2}v(\sigma+v).
    \label{explicittermLagrangian}
\end{equation}
As can be seen from Eqs.~(\ref{linearsigmamodelSSB}) and~(\ref{interactinglagrangian}) there are new terms which depend on $v$ and all fields develop dynamical masses 
\begin{align}
     m_{\sigma}^{2}&=3\lambda v^2-a^2, \nonumber \\
     m_{0}^{2}&=\lambda v^2-a^2, \nonumber \\ 
     m_{f}&=gv.
\label{masses}
\end{align}

Using Eqs.~(\ref{linearsigmamodelSSB}) and~(\ref{explicittermLagrangian}), the tree-level potential is given by
\begin{equation}
 V^{\text{tree}}(v)=-\frac{a^2+m_{0}^2}{2}v^2+\frac{\lambda}{4}v^4.
 \label{treelevel}
\end{equation}
This potential develops a minimum, called the vacuum expectation value of the $\sigma$ field, namely 
\begin{equation}
    v_{0}=\sqrt{\frac{a^2+m_{0}^2}{\lambda}}.
\label{vev}
\end{equation}
Therefore, the masses evaluated at $v_0$ are
\begin{align}
 m_f(v_0)&=g\sqrt{\frac{a^2+m_{0}^2}{\lambda}}, \nonumber \\
 m_\sigma^2(v_0)&=2a^2+3m_{0}^2, \nonumber \\
 m_{0}^2(v_0)&=m_{0}^2.
 \label{masses2}
\end{align}
Finally, an external magnetic field, uniform in space and constant in time, can be included in the model introducing a covariant derivative in the Lagrangian density, Eq.~\eqref{linearsigmamodelSSB}, namely 
\begin{equation}
 \partial_\mu\to D_\mu=\partial_\mu+iqA_\mu,
\end{equation}
where $A^\mu$ is the vector potential corresponding to an external magnetic field directed along the  $\hat{z}$ axis. In the symmetric gauge, this is given by
\begin{equation} \label{vectorpotencial}
 A^\mu(x)=\frac{1}{2}x_{\nu}F^{\nu\mu},
\end{equation}
and couples only to the charged pions  and to the quarks. 

Notice that, in order to consider the propagation of charged particles, one can resort to introduce Schwinger propagators which can be expressed either in terms of their proper time representation or as a  sum over Landau Levels. For completeness of the presentation, we now proceed to briefly discuss the properties of these propagators.


\section{\label{sec3} Magnetic field dependent boson and fermion propagators}



In order to consider the propagation of charged particles within a magnetized background, we use Schwinger's proper time representation. The fermion propagator can be written as~\cite{schwinger}
\begin{equation}
    S_f(x,x')=e^{i\Phi(x,x')}S_f(x-x'),
    \label{fermionpropagatorincoordinatespace}
\end{equation}
where $\Phi(x,x')$ is the Schwinger's phase given by
\begin{eqnarray}
\Phi(x,x')=q\int_x^{x'}d\xi_\mu \left[
A^\mu(\xi) + \frac{1}{2}F^{\mu\nu}(\xi-x')_\nu
\right],
\label{phase}
\end{eqnarray}
\begin{figure}[b]
    \includegraphics[scale=0.25]{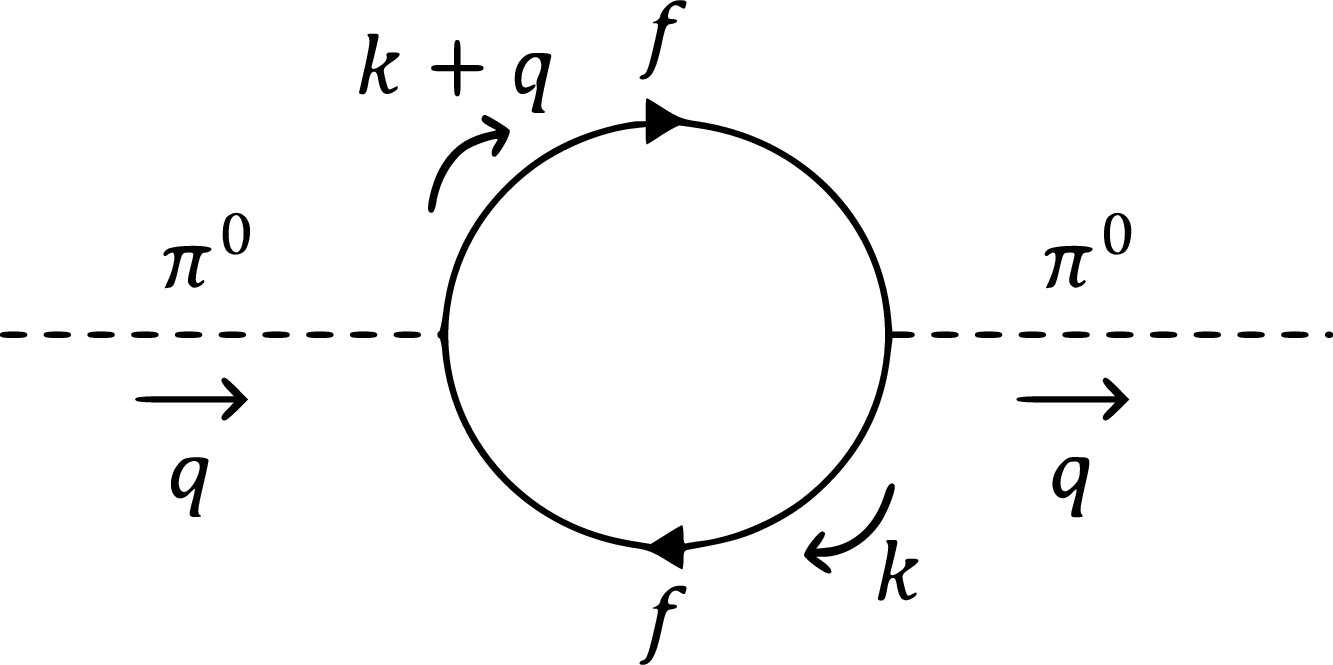}
    \caption{Feynman diagram showing the one-loop contribution from fermions to the neutral pion self-energy in the LSMq.}
    \label{self-energyquarks}
\end{figure}
where $q$ is the particle electric charge. $\Phi(x,x')$ corresponds to the translationally non-invariant and gauge dependent part of the propagator. On the other hand, $S_f(x-x')$ is translationally and gauge-invariant and can be expressed in terms of its Fourier transform as
\begin{equation}
    S_f(x-x')=\int \frac{d^{4}p}{(2\ps )^{4}}S_f(p)e^{-ip\cdot(x-x')}, \label{Fouriertransformfermionpropagator}
\end{equation}
where
\begin{eqnarray}
    iS_f(p)&=&\int_0^\infty \frac{ds}{\cos(|q_fB|s)}e^{is\left(p_\parallel^2-p_\perp^2\frac{\tan(|q_fB|s)}{|q_fB|s}-m_f^2+i\epsilon\right)}\nonumber\\
&\times&\left[
\Big(
\cos(|q_fB|s) + \gamma_1\gamma_2\sin(|q_fB|s)\text{sign}(q_fB)
\Big)\right.\nonumber\\
&\times&\left.\left(m_f +\slashed{p}_\parallel\right) - \frac{\slashed{p}_\perp}{\cos(|q_fB|s)}
\right].\label{fermionpropagatormomentumspace}
\end{eqnarray}
In a similar fashion, for a charged scalar field we have  
\begin{eqnarray}
D(x,x')&=&e^{i\Phi(x,x')}D(x-x'),\nonumber\\
D(x-x')&=&\int \frac{d^{4}p}{(2\pi)^{4}}D(p)e^{-ip\cdot(x-x')},
\label{scalarprop}
\end{eqnarray}
with
\begin{eqnarray}
iD(p)&=&\int_0^\infty \frac{ds}{\cos(|q_bB|s)}e^{is\left(p_\parallel^2-p_\perp^2\frac{\tan(|q_bB|s)}{|q_bB|s}-m_b^2+i\epsilon \right)},\nonumber\\
\label{bosonpropagatormomentumspace}
\end{eqnarray}
where the boson and fermion masses and electric charges are $m_b$, $q_b$ and $m_f$, $q_f$, respectively. 

The propagators in Eqs.~\eqref{fermionpropagatormomentumspace} and \eqref{bosonpropagatormomentumspace} can also be expanded as a sum over Landau levels. In this case, the expressions for the charged  fermion and scalar propagators are given by~\cite{Ayala-Sahu,Shovkovy}
\begin{equation}
iS_f(p)=ie^{-\frac{p_{\perp}^2}{|q_{f}B|}}\sum_{n=0}^{\infty}\frac{(-1)^{n}D_{n}(p)}{p_{\parallel}^2 -m_{f}^2 -2n|q_f B|+i\epsilon},
\label{fermionpropagatorLandauLevels}
\end{equation}
\begin{equation}
iD_b(p)=2i e^{-\frac{p_{\perp}^2}{|q_{b}B|}}\sum_{n=0}^{\infty}\frac{(-1)^{n}L_{n}^{0}\left(\frac{2p_{\perp}^{2}}{|q_{b}B|}\right)}{p_{\parallel}^2 -m_{b}^2 -(2n+1)|q_b B|+i\epsilon},
\label{bosonpropagatorLandauLevels}
\end{equation}
respectively, where
\begin{eqnarray}
D_{n}(p)&=&2(\slashed{p}_\parallel +m_f)\mathcal{O}^{+} L_{n}^{0}\left(\frac{2p_{\perp}^2}{|q_{f}B|}\right)\nonumber\\
&-&2(\slashed{p}_\parallel +m_f)\mathcal{O}^{-} L_{n-1}^{0}\left(\frac{2p_{\perp}^2}{|q_{f}B|}\right)\nonumber\\
&+&4\slashed{p}_{\perp}L_{n-1}^{1}\left(\frac{2p_{\perp}^2}{|q_{f}B|}\right),
\label{fermionfact}
\end{eqnarray}
and $L_n^m(x)$ are the generalized Laguerre polynomials. Also, in Eq.~(\ref{fermionfact}) the operators ${\mathcal{O}^{\pm}}$ are defined as
\begin{eqnarray}
{\mathcal{O}^{\pm}}=\frac{1}{2}\left( 1\pm i\gamma_1\gamma_2\, {\mbox{sign}}(qB)\right).
\label{Os}
\end{eqnarray}

We now use these ingredients to compute the elements necessary to obtain the magnetic modification of the neutral pion mass.

\section{One-loop magnetic corrections} \label{sec4}
\begin{figure}[b]
    \includegraphics[scale=0.25]{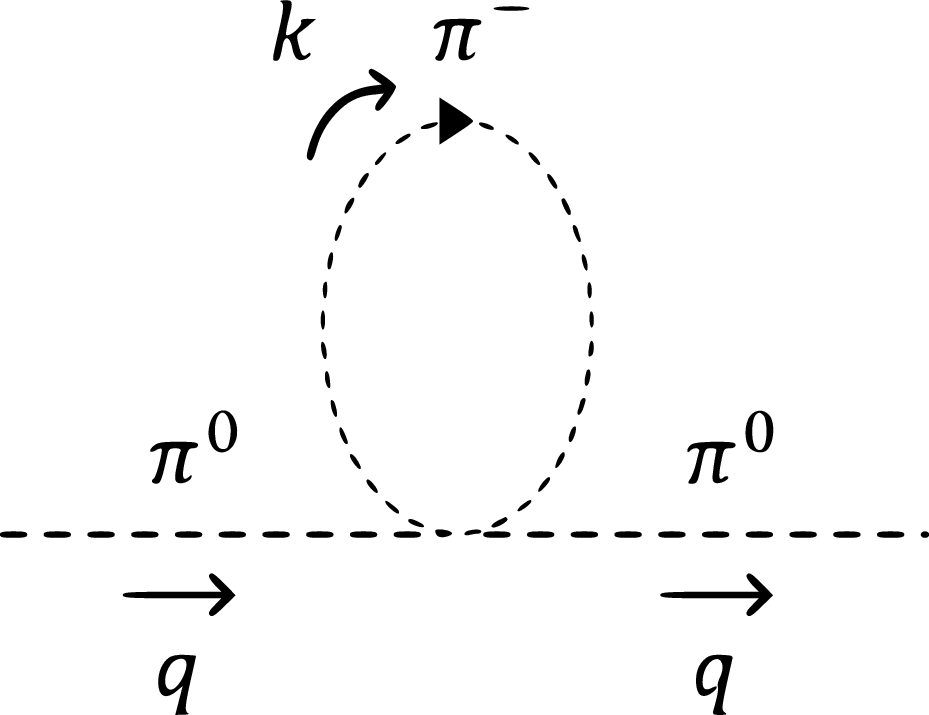}
    \caption{Feynman diagram showing the one-loop contribution from charged pions to the neutral pion self-energy in the LSMq.}
    \label{self-energypions}
\end{figure}
In order to compute the magnetic field-induced modification to the neutral pion mass, the starting point is the equation defining its dispersion relation in the presence of the magnetic field, namely
\begin{eqnarray}
  q_0^2-|\vec{q}|^2-m_0^2(B)-{\mbox{Re}}[\Pi(B,q;\lambda_B, g_B, v_B)]=0,
\label{findsol}
\end{eqnarray}
where $\Pi$ is the neutral pion self-energy and  $\lambda_B,g_B, v_B$ represent the magnetic field dependent  boson-self coupling, boson-fermion coupling and vacuum expectation value, respectively. The computation requires knowledge of each of these elements as functions of the field strength. $v_B$ can be computed finding the minimum of the magnetic field-dependent one-loop effective potential. This can be analytically computed using the full magnetic field dependence of the charged particle propagators. For  the neutral pion self-energy and the magnetic field corrections to the couplings, we work in the large field limit and thus resort to use propagators in the lowest Landau level (LLL) approximation.

\subsection{Neutral pion self-energy}
We first compute the neutral pion self-energy,
\begin{equation}
 \Pi(B,q)=\sum_f\Pi_{f\bar{f}}(B,q)+\Pi_{\pi^-}(B)+ \Pi_{\pi^+}(B)+\Pi_{\pi^0}+\Pi_{\sigma}.
 \label{totalselfenergy}
\end{equation}
The five terms on the right-hand side of Eq.~(\ref{totalselfenergy}) correspond to the Feynman diagrams contributing to this self-energy at one-loop order. The subindices represent the kind of particles in the loop. The contributions to this self-energy are: the quark-antiquark loop $\Pi_{f\bar{f}}$ depicted in Fig.~\ref{self-energyquarks} and the boson loops, $\Pi_{\pi^\pm},\Pi_{\pi^0},\Pi_{\sigma}$. The Feynman diagram corresponding to $\Pi_{\pi^-}$ is depicted in Fig.~\ref{self-energypions}, and we single it out from the neutral boson loops, since this diagram, together with the diagram corresponding to its charge conjugate (CC) $\Pi_{\pi^+}$, are the only ones modified by the presence of the magnetic field. Diagrams with neutral bosons in the loop contribute only to vacuum renormalization and not to the magnetic properties of the system. Therefore, hereafter we do not consider the latter for the description of the magnetic modifications of the pion self-energy. 

We first concentrate on the contribution from the  quark-antiquark loop for a single quark species, given explicitly by
\begin{eqnarray}
    -i\Pi_{f\bar{f}}(B,q)&=&-g^2\int \frac{d^{4}k}{(2\pi)^{4}}{\mbox{Tr}}[\gamma_{5}iS_{f}(k)\gamma_{5}iS_{f}(k+q)]\nonumber \\
    &+&\mbox{CC},
    \label{quarkloop}
\end{eqnarray}
Notice that since both particles flow with the same charge around the loop, the Schwinger's phase vanishes.
The quark propagator in the presence of a magnetic field, $iS_f$, is written in the strong field limit using the LLL contribution namely
\begin{equation}
    iS_f\to iS_f^{LLL}(k)=2ie^{\frac{-k_{\perp}^{2}}{|q_{f}B|}}\frac{\slashed{k}_{\parallel}+m_{f}}{k_{\parallel}^{2}-m_{f}^{2}+i\epsilon}\mathcal{O}^{\pm}.
    \label{fermionpropagatorLLL}
\end{equation}
According to the explicit computation in Appendix~\ref{sec7}, the fermion contribution to the pion self energy is given by
\begin{eqnarray}
    -i\Pi_{f\bar{f}}&=&\frac{ig^{2}|q_{f}B|}{2\pi^{2}}e^{-\frac{1}{2|q_{f}B|}q_{\perp}^{2}}\int_{0}^{1}dx\bigg[\frac{1}{\varepsilon}+\ln{(4\pi)}-\gamma_{E} \nonumber\\
    &-&\ln{\left( \frac{\Delta_{1}}{\mu^{2}}\right)}
    -1+\frac{x(1-x)q_{\parallel}^{2}+m_{f}^{2}}{\Delta_{1}}\bigg],
    \label{pioncont}
\end{eqnarray}
where $\Delta_{1}=x(x-1)q_{\parallel}^{2}+m_{f}^{2}$ and $\mu$ is the ultraviolet renormalization scale.

In order to capture the overall magnetic field effects for on-shell and non-moving pions, we resort to compute the fermion contribution to the pion self-energy in the
{\it static limit} namely $q_{0}=m_{B}$ and $\vec{q}=\vec{0}$. As also thoroughly discussed in Appendix~\ref{sec7}, working with the $\overline{MS}$ renormalization scheme, this is explicitly given by
\begin{eqnarray}
     \Pi_{f\bar{f}}&=&\frac{g^{2}|q_{f}B|}{2\pi^{2}}\Bigg[2\sqrt{\frac{4m_{f}^{2}}{m_B^{2}}-1}\hspace{0.1cm}\text{arccsc}\left( \frac{2m_{f}}{m_B}\right)+\ln{\left(\frac{m_{f}^{2}}{\mu^{2}}\right)}\nonumber \\
     &-&\frac{8m_{f}^{2}}{m_B\sqrt{4m_{f}^{2}-m_B^{2}}}\text{arccsc}\left(\frac{2m_{f}}{m_{B}} \right)\Bigg].
     \label{ferloop}
\end{eqnarray}
Notice that Eq.~(\ref{ferloop}) has an explicit dependence on $\mu$. This is a general feature of one loop calculations where, in order to regulate the integration, such scale needs to be introduced. As discussed in Ref.~\cite{AHHFZ}, when working in the LLL, $\mu$ needs to be chosen in such a way that this becomes the largest of all energy scales, larger than the gap $\sqrt{2|eB|}$,  between the LLL and the first excited Landau level, where $|e|$ is the absolute value of the electron charge. To accomplish this constraint, we chose $\mu^2=2|eB|+m_0^2$~\cite{AHHFZ}. With this choice, the contribution from the quark-antiquark loop for a single quark species becomes
\begin{eqnarray}
     \Pi_{f\bar{f}}&=&\frac{g^{2}|q_{f}B|}{2\pi^{2}}\Bigg[2\sqrt{\frac{4m_{f}^{2}}{m_B^{2}}-1}\hspace{0.1cm}\text{arccsc}\left( \frac{2m_{f}}{m_B}\right)\nonumber \\
     &+&\ln{\left(\frac{m_{f}^{2}}{2|eB|+m_0^2}\right)}\nonumber \\
     &-&\frac{8m_{f}^{2}}{m_B\sqrt{4m_{f}^{2}-m_B^{2}}}\text{arccsc}\left(\frac{2m_{f}}{m_{B}} \right)\Bigg].
\end{eqnarray}

We now proceed to compute the charged boson loop contribution to the pion self-energy. This can be written as 
\begin{equation}
    -i\Pi_{\pi^\pm}=\int \frac{d^{4}k}{(2\pi)^{4}}(-2i\lambda)iD_{\pi^{\pm}}(k).
    \label{chargedself}
\end{equation}
Notice that since the initial and final loop space-time points in the tadpole Feynman diagram coincide, the Schwinger's phase vanishes. To compute Eq.~(\ref{chargedself}) in the strong field limit, we use the charged boson propagator in LLL approximation, namely
\begin{equation}
    iD_{b}\to iD_{b}^{LLL}(k)=\frac{2ie^{-\frac{k_{\perp}^{2}}{|q_{b}B|}}}{k_{\parallel}^{2}-m_{b}^{2}-|q_{b}B|+i\epsilon}.
    \label{bosonpropagatorLLL}
\end{equation}
The procedure to compute this contribution is shown in Appendix~ \ref{sec7}. Choosing $\mu^2=2|eB|+m_0^2$~\cite{AHHFZ}, the result can be expressed as
\begin{equation}
   \Pi_{\pi^\pm}=-\frac{\lambda|eB| }{4\pi^{2}}\ln\left(\frac{|eB|+m_{0}^{2}}{2|eB|+m_0^2}\right).
    \label{self-energythirdcontributionLLL}
\end{equation}




With the expression for the pion self-energy at hand, we now turn our attention to compute the rest of the ingredients, starting from the magnetic corrections to the vacuum expectation value.

\subsection{Magnetic corrections to the vacuum expectation value}

The magnetic correction to the vacuum expectation value can be obtained finding the minimum for the effective potential in the presence of the magnetic background, $v_B$. For the LSMq in a magnetized medium, the effective
potential at one-loop contains fermion as well as boson contributions which modify the location of the minimum as a function of the field strength. 

The effective potential up to one-loop order has six contributions, namely
\begin{equation}
    V^{\text{eff}}=V^{\text{tree}}+V^1_{\pi^+}+V^1_{\pi^-}+V^1_{\pi^0}+V^1_{\sigma}+\sum_{f}V^1_f.
    \label{effectivepot1loop}
\end{equation}
The first term on the right-hand side of Eq.~(\ref{effectivepot1loop})  represents the classical or tree-level potential. This can be read off from Eq.~(\ref{treelevel}). The second and third terms correspond to the charged boson contribution, the fourth and fifth are the neutral contributions associated to the neutral pion and sigma, respectively, and the last one is the fermion contribution. 

The contribution to the effective potential from a charged boson with mass $m_b$ is given by the expression
\begin{equation}
V_{b}^{1}=-\frac{i}{2}\int \frac{d^{4}k}{(2\pi)^{4}}\ln\left[-D_{b}^{-1}(k)\right],
\label{Vbcont}
\end{equation}
where the charged boson propagator is given by Eq.~(\ref{bosonpropagatorLandauLevels}).
The computation of Eq.~(\ref{Vbcont}) is performed
in Appendix~\ref{sec8}. An explicit analytical expression for an arbitrary magnetic field strength can in fact be found. Working in the $\overline{MS}$ renormalization scheme and setting the boson mass to be the charged pion mass in vacuum, $m_0$, this expression is given by
\begin{eqnarray} \label{pioncargado}
V_{b}^{1}
&=& \frac{1}{16\pi^{2}}\bigg[2|eB|^{2}\psi^{-2}\left(\frac{1}{2}+\frac{m_{0}^{2}}{2|eB|}\right)-\frac{1}{2}|eB|m_{0}^{2}\ln(2\pi)\nonumber \\
&-&\frac{m_{0}^{4}}{4}\ln\left(\frac{\mu^{2}}{m_0^2} \right) - \frac{m_{0}^{4}}{4}\ln\left(\frac{m_0^{2}}{2|eB|} \right) \bigg],
\label{Vchargedboson}
\end{eqnarray}
where $\psi^{-2}(x)$ is the Polygamma function of order $-2$ and $\mu$ is the renormalization scale. Notice that in the limit $B \rightarrow 0$, 
Eq.~(\ref{Vchargedboson}) becomes
\begin{eqnarray}
V_{b}^{1}\to V_{b^0}^1
&=&-\frac{m_0^4}{64 \pi^2} \left[ \frac{3}{2} + \ln\left(\frac{\mu^{2}}{m_0^2} \right) \right],
\label{nofield}
\end{eqnarray}
which corresponds to the contribution to the effective potential from a neutral boson with mass $m_0$~\cite{villamex}. We thus use this last expression to account for the contribution coming from the fourth and fifth terms of Eq.~(\ref{effectivepot1loop}) and thus, the purely magnetic contribution from Eq.~(\ref{Vchargedboson}) is obtained by subtracting Eq.~(\ref{nofield}) from Eq.~(\ref{Vchargedboson}) and is given by
\begin{eqnarray} 
V_{b(B)}^{1}
&=& \frac{1}{16\pi^{2}}\bigg[2|eB|^{2}\psi^{-2}\left(\frac{1}{2}+\frac{m_{0}^{2}}{2|eB|}\right)-\frac{1}{2}|eB|m_{0}^{2}\ln(2\pi)\nonumber \\
&-&\frac{m_{0}^{4}}{4}\ln\left(\frac{m_0^{2}}{2|eB|} \right) + \frac{3m_0^4}{8} \bigg].
\end{eqnarray}

The contribution from a single fermion species can be obtained from the expression
\begin{equation}
V_{f}^{1}=iN_c\int \frac{d^{4}k}{(2\pi)^{4}}{\mbox{Tr}}\; \ln\left[ S_{f}^{-1}(k)\right],
\label{Vfcont}
\end{equation}
where $N_c$ is the number of colors and $iS_{f}(k)$ is given by Eqs.~(\ref{fermionpropagatorLandauLevels}) and~(\ref{fermionfact}).
The explicit computation is shown in Appendix~ \ref{sec8}. Once again, the result can be provided for an arbitrary field strength. Working with the $\overline{MS}$ renormalization scheme, this is given by
\begin{eqnarray}\label{fermion}
V_{f}^{1}&=&-\frac{N_c}{8\pi^{2}}\Biggl(4|q_{f}B|^{2}\psi^{-2}\left(\frac{m_{f}^{2}}{2|q_{f}B|}\right)-\frac{m_{f}^{4}}{2}\ln\left(\frac{\mu^{2}}{m_f^2} \right)   \nonumber \\
&-&\frac{m_{f}^{4}}{2}\ln\left(\frac{m_f^2}{2|q_{f}B|} \right) 
-m_{f}^{2}|q_{f}B|\Biggl[1+\ln(2\pi) \nonumber \\
&-&\ln\left(\frac{m_{f}^{2}}{2|q_{f}B|} \right) \Biggr] \Biggr).
\end{eqnarray}
In the limit $B \rightarrow 0$, Eq.~(\ref{fermion}) becomes
\begin{eqnarray}
V_{f}^{1}
&=&N_c\frac{m_f^4}{16 \pi^2} \left[ \frac{3}{2} + \ln\left(\frac{\mu^{2}}{m_f^2} \right) \right],
\end{eqnarray}
which corresponds to the contribution to the effective potential from a fermion in the absence of the magnetic field. Thus the purely magnetic contribution from Eq.~(\ref{fermion}) is given by
\begin{eqnarray}
V_{f(B)}^{1}&=&-\frac{N_c}{8\pi^{2}}\Biggl(4|q_{f}B|^{2}\psi^{-2}\left(\frac{m_{f}^{2}}{2|q_{f}B|}\right) \nonumber \\
&-&\frac{m_{f}^{4}}{2}\ln\left(\frac{m_f^2}{2|q_{f}B|}\right)- 
m_{f}^{2}|q_{f}B|\Biggl[1+\ln(2\pi) \nonumber \\
&-&\ln\left(\frac{m_{f}^{2}}{2|q_{f}B|} \right) \Biggr] + \frac{3m_f^4}{4} \Biggr) .
\end{eqnarray}
\begin{figure}[t]
    \centering
    \includegraphics[scale=0.59]{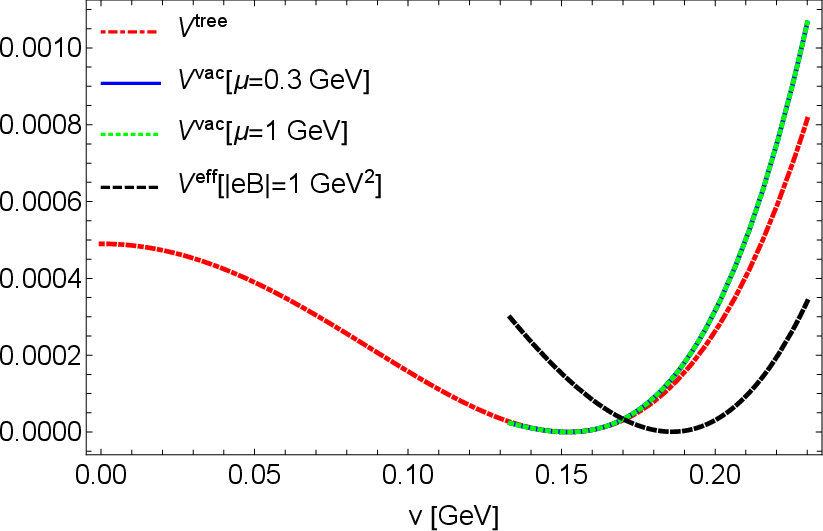}
    \caption{Comparison between the position and curvature of the minimum of $V^{\text{tree}}$ and $V^{\text{vac}}$ computed with $\mu=0.3,1$ GeV, after implementing the vacuum stability conditions. Also shown an example of the position of the minimum for $V^{\text{eff}}$, $v_B$, computed with $|eB|=1$ GeV$^2$. For the calculation we use $m_0=140$ MeV, $\lambda=3.67$, $g=0.46$, and correspondingly $m_\sigma=435$ MeV, $a=256$ MeV and $v_0=152$ MeV.}
    \label{fig3}
\end{figure}
When the tree-level effective potential is modified by one-loop corrections, the curvature (or equivalently, the vacuum $\sigma$ mass) and the position of the minimum are bound to change. The changes are driven both from purely vacuum contributions as well as from magnetic field effects. The vacuum changes need to be absorbed with a redefinition of the vacuum terms so as to make sure that any change in the position of the minimum truly comes from the magnetized background. This is accomplished by enforcing the {\it vacuum stability conditions}~\cite{RMF}, introducing counterterms in such a way that 
\begin{eqnarray}
V^{\text{tree}} &=&-\frac{(a^2+m_0^2)}{2}v^2 + \frac{\lambda}{4}v^4 \nonumber \\
&\to&\nonumber\\
V^{\text{tree}}+\delta V^{\text{tree}} &=& - \frac{(a^2+m_0^2+\delta a^2)}{2}v^2 + \frac{(\lambda+\delta \lambda)}{4}v^4, \nonumber \\
\end{eqnarray}
where $\delta a^2$ and $\delta\lambda$ are to be determined from the conditions
\begin{align}\label{v02}
\frac{1}{2v}\frac{dV^{\text{vac}}}{dv} \Big|_{v=v_0}&=0,\nonumber \\
\frac{d^2V^{\text{vac}}}{dv^2} \Big|_{v=v_0}&=2a^2+2m_0^2.
\end{align}
$V^{\text{vac}}$ contains the contribution from the three pions, the $\sigma$ and the three color charges for the two light quarks, in the limit $B\to 0$, namely 
\begin{eqnarray}
&&V^{\text{vac}}=- \frac{(a^2+m_0^2+\delta a^2)}{2}v^2 + \frac{(\lambda+\delta \lambda)}{4}v^4 \nonumber \\
&-&3\frac{m_0^4}{64 \pi^2} \left[ \frac{3}{2} + \ln\left(\frac{\mu^{2}}{m_0^2} \right) \right] - \frac{m_{\sigma}^4}{64 \pi^2} \left[ \frac{3}{2} +\ln\left(\frac{\mu^{2}}{m_{\sigma}^2} \right) \right] \nonumber \\&+&2 N_c\frac{m_f^4}{16 \pi^2} \left[ \frac{3}{2} + \ln\left(\frac{\mu^{2}}{m_f^2} \right) \right].
\end{eqnarray}

\begin{figure}[t]
    \centering
    \includegraphics[scale=0.59]{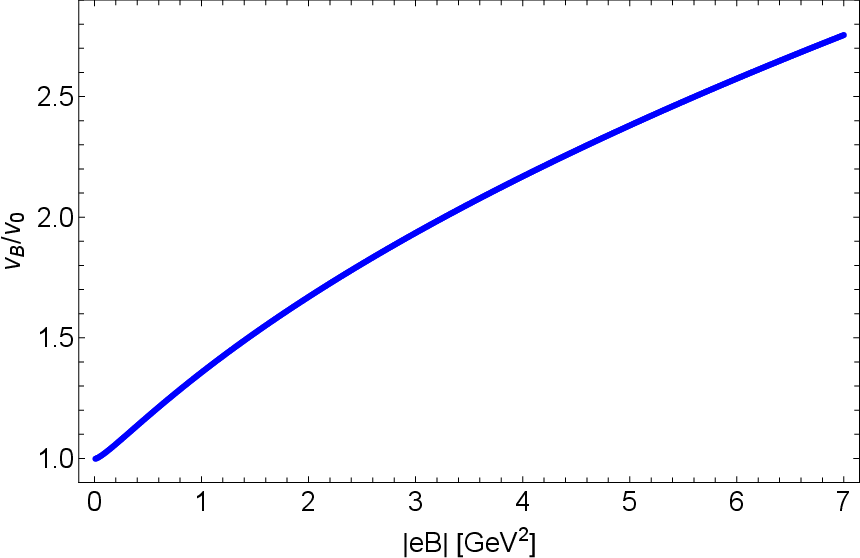}
    \caption{Magnetic modification to the vacuum expectation value, $v_B$, as a function of the field strength. For the calculation we use $m_0=140$ MeV, $\lambda=3.67$, $g=0.46$, and correspondingly $m_\sigma=435$ MeV, $a=256$ MeV and $v_0=152$ MeV.}
    \label{fig4}
\end{figure}

With this procedure we obtain  
\begin{eqnarray}
  &&  \delta a^2=\frac{1}{16\pi^2 \lambda} \Biggl[8 a^2 g^4 N_c + 8g^4 m_0^2 N_c - 6 a^2 \lambda^2   \nonumber \\
    &-&12 m_0^2 \lambda^2 + 3 a^2 \lambda^2 \ln\left(\frac{\mu^{2}}{m_0^2} \right) + 3 a^2 \lambda^2  \ln\left(\frac{\mu^{2}}{2 a^2+3 m_0^2} \right)  \Biggr], \nonumber \\
\end{eqnarray}
\begin{eqnarray}
&&\delta \lambda = \frac{1}{16\pi^2} \Biggl[  3\lambda^2  \ln\left(\frac{\mu^{2}}{m_0^2} \right) + 9  \lambda^2  \ln\left(\frac{\mu^{2}}{2 a^2+3 m_0^2} \right)\nonumber \\
&-&8g^4 N_c  \ln\left(\frac{\lambda \mu^{2}}{g^2(a^2+m_0^2)} \right)\Biggr].
\end{eqnarray}
Thus, once the vacuum terms --evaluated at the vacuum expectation value-- are included into the effective potential, the modifications to the minimum come exclusively from magnetic effects, namely, from the contribution of charged pions and fermions. As a result, the one-loop effective potential in a magnetized medium can be written as
\begin{eqnarray}
&V^{eff}(B)&=- \frac{(a^2+m_{0}^2)}{2}v^2-\frac{\delta a^2}{2}v_0^2 + \frac{\lambda}{4}v^4+\frac{\delta \lambda}{4}v_0^4 \nonumber \\
&-&3\frac{m_{0}^4(v_0)}{64 \pi^2} \Biggl[ \frac{3}{2} + \ln\left(\frac{\mu^{2}}{m_{0}^2(v_0)} \right) \Biggr]   \nonumber \\
&-&\frac{m_{\sigma}^4(v_0)}{64 \pi^2} \; \Biggl[ \frac{3}{2}+\ln\left(\frac{\mu^{2}}{m_{\sigma}^2(v_0)} \right) \Biggr]\nonumber \\
&+&2 N_c\sum_{f}\frac{m_f^4(v_0)}{16 \pi^2} \; \Biggl[ \frac{3}{2} + \ln\left(\frac{\mu^{2}}{m_f^2(v_0)} \right) \Biggr]\nonumber \\
&+&\frac{2}{16\pi^{2}}\;\Biggl[2|eB|^{2}\,\psi^{-2}\left(\frac{1}{2}+\frac{m_{0}^{2}(v)}{2|eB|}\right) + \frac{3m_0^4(v)}{8} \nonumber \\
&-&\frac{1}{2}|eB|\,m_{0}^{2}(v)\ln(2\pi)-\frac{m_{0}^{4}(v)}{4}\ln\left(\frac{m_0^{2}(v)}{2|eB|} \right) \Biggr] \nonumber \\
&-&\frac{N_c}{8\pi^{2}}\sum_{f}\Bigg[4|q_{f}B|^{2}\psi^{-2}\left(\frac{m_{f}^{2}(v)}{2|q_{f}B|}\right)+\frac{3}{4} m_{f}^{4}(v) \nonumber \\
&-&\frac{m_{f}^{4}(v)}{2} \ln\left(\frac{m_f^2(v)}{2|q_{f}B|}\right)-m_{f}^{2}(v)|q_{f}B|\nonumber \\ &+&m_{f}^{2}(v)|q_{f}B|\ln\left(\frac{m_{f}^{2}(v)}{4\pi|q_{f}B|} \right)\Bigg].
\end{eqnarray}
Figure~\ref{fig3} shows the tree-level potential $V^{\text{tree}}$ and the vacuum one-loop potential $V^{\text{vac}}$ computed for  $\mu=0.3, 1$ GeV, after implementing the stability conditions. Also shown in the figure is the magnetic field-modified position of the minimum when adding the magnetic effects to $V^{\text{vac}}$, for $|eB|=1$ GeV$^2$. Notice that after the vacuum stability conditions are implemented, the vacuum position and curvature remain at their tree-level values and that these quantities are independent of the choice of the renormalization scale $\mu$. Figure~\ref{fig4} shows the position of the minimum, $v_B$, as a function of the field strength. Notice that, as expected, $v_B$ grows with the field strength, signaling  magnetic catalysis. 

\subsection{Magnetic modifications to the boson self-coupling and boson-fermion coupling}
\begin{figure}[t]
    \centering
    \includegraphics[scale=0.59]{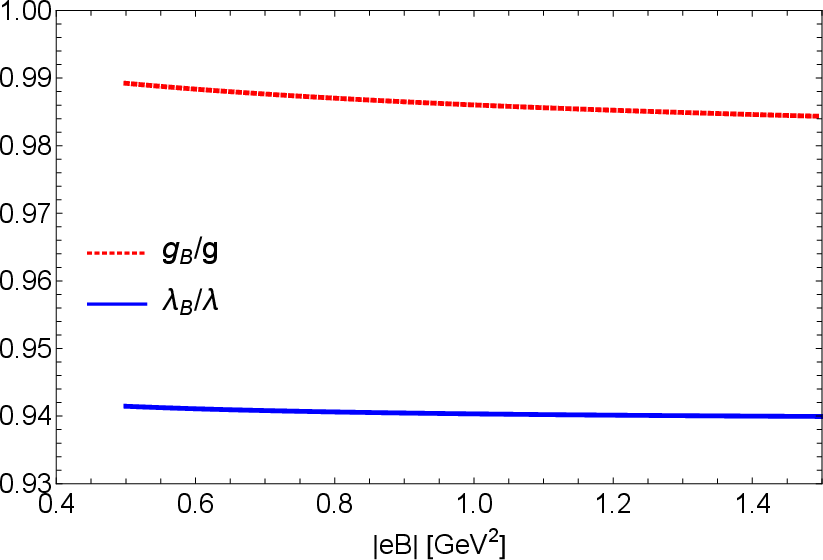}
    \caption{Magnetic modification to the boson-fermion coupling (dotted red line) and to the boson self-coupling (solid blue line) as a function of the field strength. For the calculation we use $m_0=140$ MeV, $\lambda=3.67$, $g=0.46$, and correspondingly $m_\sigma=435$ MeV, $a=256$ MeV and $v_0=152$ MeV.}
    \label{fig5}
\end{figure}

The magnetic field-induced corrections to the boson self-coupling $\lambda$ and the boson-fermion coupling $g$ have been recently obtained in Ref.~\cite{AHHFZ}. Working in the strong field limit, the explicit expressions for these corrections are given by
\begin{equation}
    \Gamma_{\lambda}^{LLL}=-\frac{\lambda}{6\pi^{2}}  \frac{|q_{b}B|}{|q_{b}B|+m_{0}^{2}},
    \label{magneticcorrectiontolambdaLLLallmomentazero}
\end{equation}
and
\begin{equation}
    \Gamma_{g}^{LLL}=\Gamma^{LLL}_{1,g}+\Gamma^{LLL}_{2,g}+\Gamma^{LLL}_{3,g},
\end{equation}
where
\begin{eqnarray}
    \Gamma_{1,g}^{LLL}&=&\frac{g^{2}|eB|}{16\pi^{2}m_{f}^{2}}\int_{0}^{1}du\frac{u}{ u^{2}+\alpha(1-u)}\nonumber\\
    &\times&\left[1+\frac{(2-u)u}{u^{2}+\alpha(1-u)} \right],\nonumber\\
   \Gamma_{2,g}^{LLL}&=&-\frac{g^{2}}{2\pi^{2} m_{f}^{2}}\int_{0}^{1}du \int_{0}^{\infty} dk_{\perp}\; k_{\perp} e^{-\frac{3k_{\perp}^{2}}{|eB|}} \nonumber\\
   &\times& \frac{u}{u^{2}+\beta(1-u)}\left[1+\frac{(2-u)u}{ u^{2}+\beta(1-u)} \right],\nonumber\\
   \Gamma_{3,g}^{LLL}&=&\frac{g^{2}}{2\pi^{2}m_{f}^{2}}\int_{0}^{1}du \int_{0}^{\infty} dk_{\perp}\;
    k_{\perp}e^{-\frac{3k_{\perp}^{2}}{|eB|}} \nonumber\\
    &\times&\frac{u}{u^{2}+\gamma(1-u)}\left[1+\frac{(2-u)u}{u^{2}+\gamma(1-u)} \right],
    \label{resultGammag3}   
\end{eqnarray}
with $\alpha=(m_{0}^{2}+|eB|)/m_{f}^{2}$, $\beta= (k_{\perp}^{2}+m_{0}^{2})/m_{f}^{2}$ and $\gamma=(k_{\perp}^{2}+m_{\sigma}^{2})/m_{f}^{2}$. 
\begin{figure}[t]
    \centering
    \includegraphics[scale=0.59]{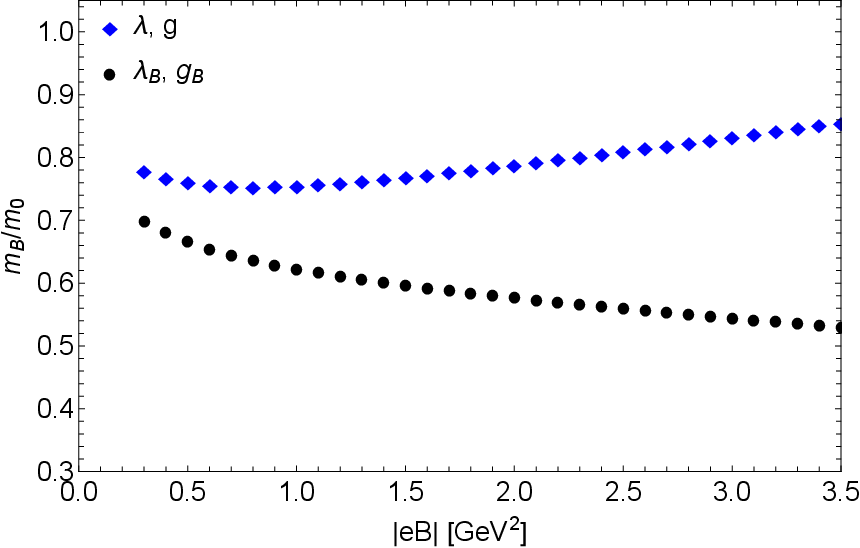}
    \caption{Magnetic modification to the neutral pion mass, for two different cases: blue diamonds correspond to tree level coupling couplings and black dots correspond to magnetic field-dependent couplings. Notice that, whereas the calculation using tree-level couplings starts off decreasing to later on increase, the calculation using magnetic field-dependent couplings decreases monotonically as a function of the field strength. For the calculation we use $m_0=140$ MeV, $\lambda=3.67$, $g=0.46$, and correspondingly $m_\sigma=435$ MeV, $a=256$ MeV and $v_0=152$ MeV.}
    \label{fig6}
\end{figure}

The magnetic modified boson self-coupling and boson-fermion coupling are thus given by
\begin{eqnarray}
    \lambda_B&=&\lambda\left(1 + \Gamma_\lambda^{LLL}\right),\nonumber\\
    g_B&=&g\left(1+\Gamma_{g}^{LLL} \right).\label{couplingsN}
\end{eqnarray}
Figure~\ref{fig5} shows the magnetic field dependence of these couplings, normalized to their vacuum values. Notice that the couplings show a monotonic, albeit modest, decrease with the field strength. 

\section{Magnetic modification to the neutral pion mass} \label{sec5}

With all the elements at hand, we can now find the magnetic field-dependent neutral pion mass from the dispersion relation, Eq.~(\ref{findsol}), in the limit where $\vec{q}\to 0$ and $q_0\to m_B$, namely 
\begin{equation}
m_B^{2}=m_{0}^{2}(B)+\Pi(B,q_0=m_B,\vec{q}=0;\lambda_B,g_B,v_B),
\label{selfconsistenteq}
\end{equation}
where, in order to incorporate the magnetic field-dependent boson self-coupling and vacuum expectation value in the three-level pion mass, we write 
\bea
m_{0}^{2}(B)=\lambda_B v_B^{2}-a^{2}.
\label{threelevelmod}
\eea
\begin{figure}[t]
    \centering
    \includegraphics[scale=0.59]{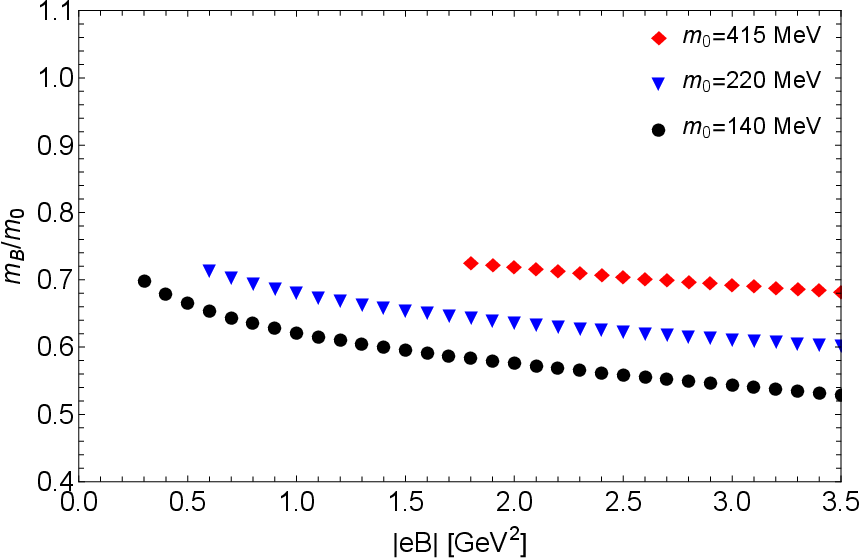}
    \caption{Magnetic modification to the neutral pion mass for three different values of pion mass in vacuum: black dots, $m_0=140$ MeV, blue triangles, $m_0=220$ MeV and red diamonds $m_0=415$ MeV. Notice that, as the vacuum pion mass decreases the corresponding magnetic field-dependent pion mass also decreases and that all cases show a monotonic decrease as a function of the field strength, in agreement with the recent LQCD findings.}
    \label{fig7}
\end{figure}
To reduce the parameter space, we consider that in the absence of baryons, the constituent quark mass is such that $m_0=2m_f$. With this choice, the only free parameters are $\lambda$ and  $g$. We have explored a large range for these parameters and hereby we show the results for the set that best describes simultaneously the LQCD data of Refs.~\cite{Endrodi,Ding}. Moreover, the values we use as initial inputs for $g$ and $\lambda$ produce, for the lowest vacuum pion mass considered, values of $v_0$ larger than $f_\pi$ only by a factor $\sim 1.5$, which we take as an indication of consistency within the limitations of an effective theory such as the LSMq. Since Refs.~\cite{Endrodi,Ding} report their findings for different values of the vacuum pion mass, we also vary this mass and consequently the rest of the dependent parameters have to be changed to suit these choices. In particular, a larger vacuum pion mass implies a larger $\sigma$ mass. Thus, in the strong field limit, our results are restricted to the domain where $|eB|>m_\sigma^2$.

\begin{figure}[b]
    \centering
    \includegraphics[scale=0.59]{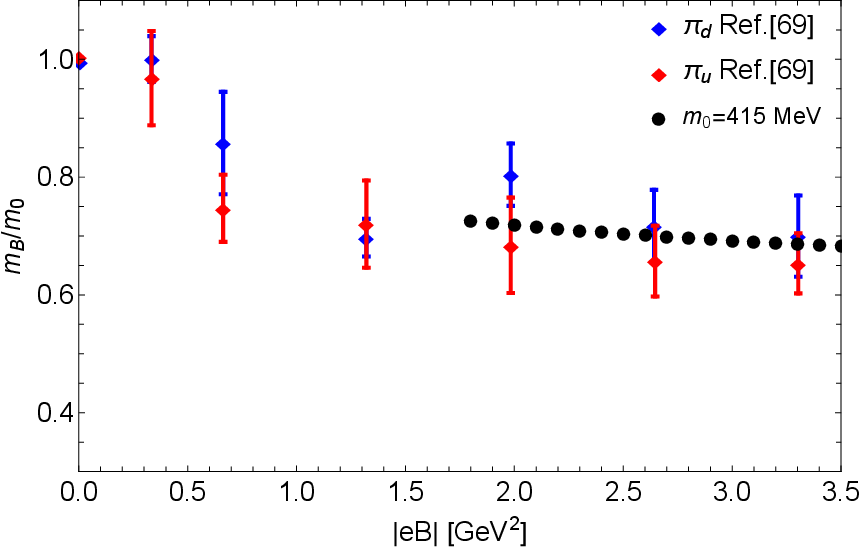}
    \caption{Magnetic modification to the neutral pion mass. Blue and red diamonds correspond to the masses of $\pi_d$ and $\pi_u$ reported by LQCD in Ref.~\cite{Endrodi} with $m_0=415$ MeV. Black dots are the result from Eq.~(\ref{selfconsistenteq}) with $m_0=415$ MeV, $\lambda=3.67$, $g=0.46$, and correspondingly $m_\sigma=1291$ MeV, $a=758$ MeV and $v_0=451$ MeV.}
    \label{fig8}
\end{figure}
Figure~\ref{fig6} shows the magnetic field-dependent neutral pion mass as a function of the field strength computed for two cases: with (black dots) and without (blue diamonds) magnetic field-dependent couplings, using as inputs $m_0=140$ MeV, $\lambda=3.67$, $g=0.46$, and correspondingly $m_\sigma=435$ MeV, $a=256$ MeV and $v_0=152$ MeV. Notice that, whereas the former shows a monotonic decrease, the latter starts off decreasing to later on increase as a function of the field strength. This result signals the importance of including magnetic field corrections to the couplings in the calculation of the magnetic field-dependent neutral pion mass.

In order to compare with LQCD simulations, which are implemented for different values of the vacuum pion mass, Fig.~ \ref{fig7} shows the magnetic field-dependent neutral pion mass as a function of the field strength when varying the input vacuum pion mass. Shown are three cases: $m_0=140$ MeV (black dots),  $m_0=220$ MeV (blue triangles) and $m_0=415$ MeV (red diamonds). Notice that, as the vacuum pion mass decreases, the corresponding magnetic field-dependent pion mass also decreases and that all cases show a monotonic decrease as a function of the field strength, in agreement with the LQCD findings.
\begin{figure}[t]
    \centering
    \includegraphics[scale=0.59]{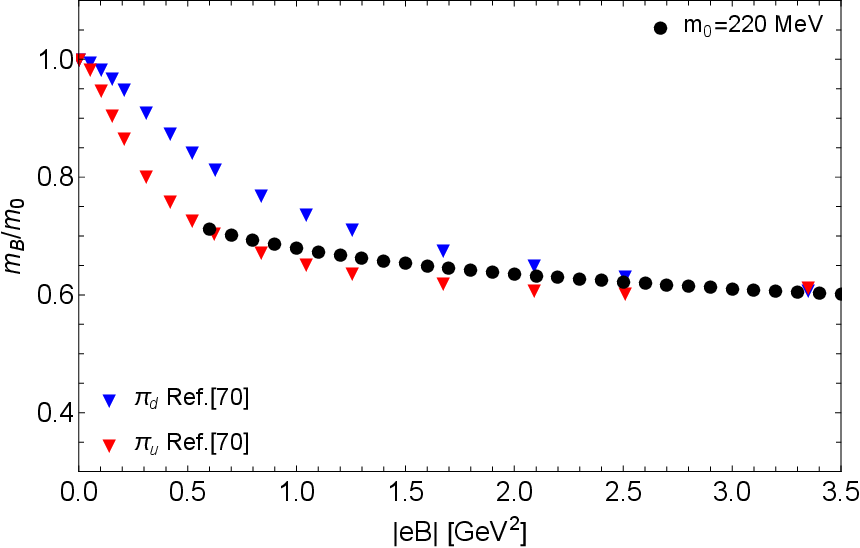}
    \caption{Magnetic modification to the neutral pion mass. Blue and red diamonds correspond to the masses of $\pi_d$ and $\pi_u$ reported by LQCD in Ref.~\cite{Ding} with $m_0=220$ MeV. Black dots are the result from Eq.~(\ref{selfconsistenteq}) with $m_0=220$ MeV, $\lambda=3.67$, $g=0.46$, and correspondingly $m_\sigma=684$ MeV, $a=402$ MeV and $v_0=239$ MeV.}
    \label{fig9}
\end{figure}
\begin{figure}[b]
    \centering
    \includegraphics[scale=0.59]{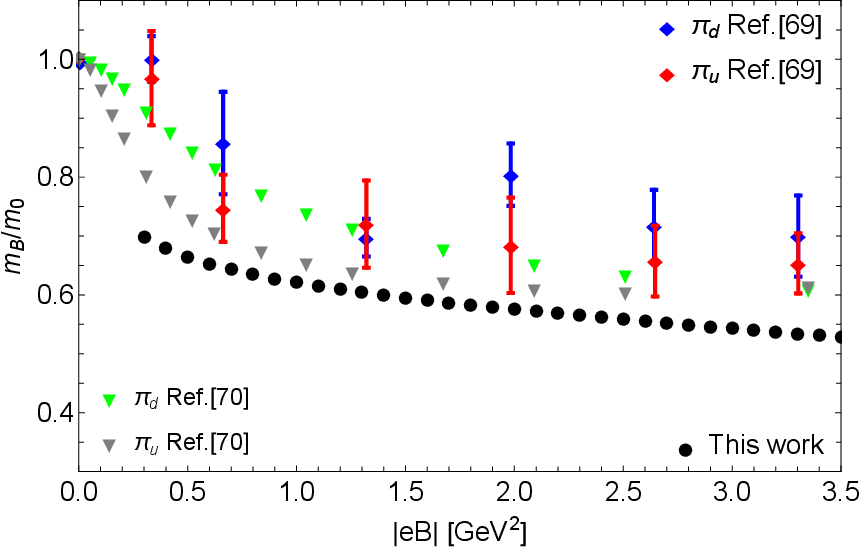}
    \caption{Magnetic modification to the neutral pion mass. Blue and red diamonds correspond to the masses of $\pi_d$ and $\pi_u$ reported by LQCD in Ref.~\cite{Endrodi} with $m_0=415$ MeV. Green and gray triangles correspond to the masses of $\pi_d$ and $\pi_u$ reported by LQCD in Ref.~\cite{Ding} with $m_0=220$ MeV. Black dots are the result of Eq.~(\ref{selfconsistenteq}) with $m_0=140$ MeV, $\lambda=3.67$, $g=0.46$, and correspondingly $m_\sigma=435$ MeV, $a=256$ MeV and $v_0=152$ MeV. Notice that, as expected, when for the calculation we use as input the physical pion mass $m_0 = 140$ MeV, the theoretical curve lies below the LQCD data which were obtained using larger vacuum pion masses.}
    \label{fig10}
\end{figure}
To make direct contact with LQCD data, Fig.~\ref{fig8} shows the results for the magnetic field-dependent neutral pion mass as a function of the field strength using as input $m_0=415$ MeV and with $\lambda=3.67$ and $g=0.46$, compared to the results from Ref.~\cite{Endrodi}. The data points correspond to the $\pi_d$ (blue diamonds) and $\pi_u$ (red diamonds) masses computed also using as input $m_0=415$ MeV. Notice that our calculation does a nice description of the data average, particularly for the largest field strengths. Figure~\ref{fig9} shows also a comparison of our calculation with the LQCD calculation of Ref.~\cite{Ding}, this time computed with $m_0=220$ MeV as input together with  $\lambda=3.67$ and $g=0.46$, The data points correspond to the $\pi_d$ (blue diamonds) and $\pi_u$ (red diamonds) masses computed also using as input  $m_0=220$ MeV. Once again, we notice that our calculation does a nice job describing the average of the LQCD masses, particularly for large values of the field strength.

Finally, Fig.~\ref{fig10} shows a comparison of our calculation with the results of the LQCD calculations from Refs.~\cite{Endrodi} and~\cite{Ding}. The data points correspond to the lowest reached values of each LQCD calculation, $m_0=415$ MeV for the former and $m_0=220$ MeV for the latter. The calculation (black dots) is performed with $m_0=140$ MeV as input, together with  $\lambda=3.67$ and $g=0.46$. Notice that the result of the calculation using as input the physical pion mass in vacuum lies below the LQCD points. In this sense, this result can be considered as our prediction when and if LQCD techniques can be performed for a physical vacuum pion mass.

\section{Summary and conclusions}\label{sec6}

In this work we have used the LSMq to find the magnetic field-induced modifications to the neutral pion mass at one-loop level. The magnetic field effects are introduced by using charged particle propagators in the presence of a magnetic background in the strong field limit. We found that the approach is able to reproduce the qualitative and quantitative magnetic field dependence of the neutral pion mass reported by recent LQCD calculations. The important ingredients for the calculation are the proper inclusion of the magnetic field effects on the model couplings, the $\sigma$ vacuum expectation value and the neutral pion self-energy. As shown, the magnetic field effects produce that the model couplings monotonically decrease as a function of the field strength. When this behavior is not accounted for, we have shown that the neutral pion mass starts off decreasing to then increase at an intermediate value of the field strength. We have also shown that by accounting for the vacuum stability conditions, the magnetic field-dependent vacuum expectation value of the $\sigma$-field increases as a function of the field strength, which is to be expected on general grounds given the well established magnetic catalysis phenomenon of the condensate in a magnetized medium. This increase is however outdone by the behaviour of the  pion self-energy as a function of the field strength such that overall, the neutral pion mass monotonically decreases as a function of the field strength. By comparing to the LQCD calculations performed for the smallest pion mass allowed by that technique, we show that when the physical vacuum pion mass is used, the magnetic field-dependent neutral pion mass curve lies a bit below the LQCD data. In this sense, this result is our prediction for when and if the LQCD techniques allow for calculations using the physical vacuum pion mass.

An interesting question is whether our approach can also reproduce the magnetic field behavior of the charged pion mass as well as that of the mass of other mesons such as the neutral and charged $\rho$ mesons. This is work in progress and will be reported elsewhere.

\section*{Acknowledgements}
This work was supported in part by UNAM-DGAPA-PAPIIT grant number IG100219 and by Consejo Nacional de Ciencia y Tecnolog\'{\i}a grant numbers A1-S-7655 and A1-S-16215. L. A. H. acknowledges support from a PAPIIT-DGAPA-UNAM fellowship. R. Zamora acknowledges support from FONDECYT (Chile) under grant No. 1200483. This work is partially supported by Conselho Nacional de Desenvolvimento Cient\'ifico e Tecnol\'ogico  (CNPq), Grant No. 304758/2017-5 (R. L. S. F); Funda\c{c}\~ao de Amparo \`a Pesquisa do Estado do Rio Grande do Sul (FAPERGS), Grants Nos. 19/2551- 0000690-0 and 19/2551-0001948-3 (R. L. S. F.).
\begin{appendix}
\section{Magnetic corrections to the neutral pion self-energy}\label{sec7}
Consider the quark loop which can be made either of quarks $u$ or $d$, as depicted in Fig.~\ref{self-energyquarks}. The contribution from a quark flavor $f$ is given by
\begin{eqnarray}
    -i\Pi_{f\bar{f}}(B,q)&=&-g^2\int \frac{d^{4}k}{(2\pi)^{4}}Tr[\gamma^{5}iS_{f}(k)\gamma^{5}iS_{f}(k+q)]\nonumber \\
    &+&\mbox{CC},
\end{eqnarray}
where we used that the Schwinger's phase vanishes.
We now use Eq.~\eqref{fermionpropagatorLLL} to account for the strong field limit, and the properties of the Dirac matrices
\begin{eqnarray}
    &&\mathcal{O}^{\pm}\slashed{a}_{\parallel}=\slashed{a}_{\parallel}\mathcal{O}^{\pm}, \nonumber \\ 
    &&\mathcal{O}^{\pm}\gamma^{5}=\gamma^{5}\mathcal{O}^{\pm},\nonumber \\
    &&\mathcal{O}^{+}+\mathcal{O}^{-}=\mathcal{I}_{4\times4}, \nonumber \\
    &&\left(\mathcal{O}^{\pm}\right)^{2}=\mathcal{O}^{\pm}, \nonumber \\
    &&\gamma^{5}\slashed{a}_{\parallel}=-\slashed{a}_{\parallel}\gamma^{5},  
    \label{gammaproperties}
\end{eqnarray}
where $a^{\mu}_{\parallel}=(a_0,0,0,a_3)$ and $\slashed{a}_{\parallel}=a_{\parallel\mu}\gamma^{\mu}$. Adding up the contribution from the CC diagram
we get
\begin{equation}
    -i\Pi_{f\bar{f}}=4g^2\int \frac{d^{4}k}{(2\pi)^{4}}e^{-\frac{k_{\perp}^{2}+(k+q)_{\perp}^{2}}{|q_{f}B|}}\frac{\mathcal{N}}{AB}, 
\end{equation}
where we define 
\begin{eqnarray}
    &&\mathcal{N}\equiv Tr\left[(m_{f}-\slashed{k}_{\parallel})((\slashed{k}+\slashed{q})_{\parallel}+m_{f})\right] \nonumber \\
    &&A=(k+q)_{\parallel}^{2}-m_{f}^{2}+i\epsilon, \nonumber \\
    &&B=k_{\parallel}^{2}-m_{f}^{2}+i\epsilon.
\end{eqnarray}
We proceed to integrate  over  the  perpendicular coordinates relative to the magnetic field.  The result is given by
\begin{equation}
    -i\Pi_{f\bar{f}}=g^{2}\frac{|q_{f}B|}{2\pi}e^{-\frac{1}{2|q_{f}B|}q_{\perp}^{2}}\int \frac{d^{2}k_{\parallel}}{(2\pi)^{2}}\frac{\mathcal{N}}{AB}.
    \label{quarkloopexp1}
\end{equation}
We introduce the Feynman parametrization
\begin{equation}
    \frac{1}{AB}=\int_{0}^{1}\frac{dx}{\left[Ax+B(1-x) \right]^{2}}.
\end{equation}
The denominator of Eq.~\eqref{quarkloopexp1} can be written as
\begin{equation}
    Ax+B(1-x)=(k+xq_{\parallel})^{2}-\Delta_{1}+i\epsilon,
\end{equation}
where $\Delta_{1}=x(x-1)q_{\parallel}^{2}+m_{f}^{2}$ and $\epsilon \to 0$. We make the change of variables $k_{\parallel}=l_{\parallel}-xq_{\parallel}$ such that the numerator, $\mathcal{N}$, can be expressed as
\begin{equation}
    \mathcal{N}=4m_{f}^{2}-4l_{\parallel}^{2}+4x(1-x)q_{\parallel}^{2}.
\end{equation}
Notice that we have taken into account that the trace of an odd number of Dirac matrices vanishes and that the linear term $l_{\parallel}$ will vanish in the integration. Thus, the contribution to the self-energy becomes
\begin{eqnarray}
    -i\Pi_{f\bar{f}}&=&\frac{2g^{2}|q_{f}B|}{\pi}e^{-\frac{1}{2|q_{f}B|}q_{\perp}^{2}}\int_{0}^{1}dx\int \frac{d^{2}l_{\parallel}}{(2\pi)^{2}}\bigg[\frac{-l_{\parallel}^{2}}{(l_{\parallel}^{2}-\Delta_{1})^{2}} \nonumber\\
    &+&\frac{x(1-x)q_{\parallel}^{2}+m_{f}^{2}}{(l_{\parallel}^{2}-\Delta_{1})^{2}} \bigg].
\end{eqnarray}
In order to find the integral over parallel coordinates relative to the magnetic field we proceed using dimensional regularization
\begin{equation}
    \int\frac{d^{4}k_{E}}{(2\pi)^{4}}\rightarrow \mu^{4-d}\int\frac{d^{d-2}k_{E\parallel}}{(2\pi)^{d-2}}\int\frac{d^{2}k_{\perp}}{(2\pi)^{2}},
    \label{dimensionalreg}
\end{equation}
where we use $d=4-2\varepsilon$.
In order to perform the integral over $d^{2}l_{\parallel}$ we use that
\begin{equation}
    \mu^{4-d}\int\frac{d^{d-2}l_{\parallel}}{(2\pi)^{d-2}} \frac{1}{(l_{\parallel}^{2}-\Delta)^{2}}=\frac{i}{4\pi \Delta}\left[1+\mathcal{O}(\varepsilon)\right],
    \label{dimensionalregularizationparallel1}
\end{equation}
\begin{eqnarray}
    \mu^{4-d}\int\frac{d^{d-2}l_{\parallel}}{(2\pi)^{d-2}} \frac{l_{\parallel}^{2}}{(l_{\parallel}^{2}-\Delta)^{2}}&=&-\frac{i}{4\pi}\bigg[\frac{1}{\varepsilon}+\ln{(4\pi)}-\gamma_{E}-1 \nonumber\\
    &-&\ln{\left( \frac{\Delta}{\mu^{2}}\right)}+\mathcal{O}(\varepsilon)\bigg].
    \label{dimensionalregularizationparallel2}
\end{eqnarray}
According to Eqs.~\eqref{dimensionalregularizationparallel1} and~\eqref{dimensionalregularizationparallel2}
\begin{eqnarray}
    -i\Pi_{f\bar{f}}&=&\frac{ig^{2}|q_{f}B|}{2\pi^{2}}e^{-\frac{1}{2|q_{f}B|}q_{\perp}^{2}}\int_{0}^{1}dx\bigg[\frac{1}{\varepsilon}+\ln{(4\pi)}-\gamma_{E} \nonumber\\
    &-&\ln{\left( \frac{\Delta}{\mu^{2}}\right)}
    -1+\frac{x(1-x)q_{\parallel}^{2}+m_{f}^{2}}{\Delta}\bigg].
\end{eqnarray}
In the static limit, $\vec{q}=\vec{0}$, and setting the zeroth component of the momentum equal to the neutral pion mass, $q_{0}=m_{B}$ one can solve  Eq.~\eqref{selfconsistenteq} self-consistently. 
We proceed using the $\overline{MS}$ renormalization scheme to obtain a finite expression given by
\begin{eqnarray}
    -i\Pi_{f\bar{f}}&=&\frac{ig^{2}|q_{f}B|}{2\pi^{2}}\int_{0}^{1}dx\bigg[\frac{m_{f}^{2}-x(x-1)m_{B}^{2}}{m_{f}^{2}+x(x-1)m_{B}^{2}}-1 \nonumber\\
    &-&\ln{\left(\frac{x(x-1)m_{B}^{2}+m_{f}^{2}}{\mu^{2}}\right)} \bigg].
\end{eqnarray}
The integration over the Feynman parameter can be performed provided that $4m_{f}^{2}>m_{B}^{2}$, this conditions is the threshold relation for this process and it must remain valid upon the choice of the set of parameters. Substituting and reducing terms, we get
\begin{eqnarray}
     \Pi_{f\bar{f}}&=&\frac{g^{2}|q_{f}B|}{2\pi^{2}}\Bigg[2\sqrt{\frac{4m_{f}^{2}}{m_{B}^{2}}-1}\hspace{0.1cm}\text{arccsc}\left( \frac{2m_{f}}{M_{B}}\right)+\ln{\left(\frac{m_{f}^{2}}{\mu^{2}}\right)}\nonumber \\
     &-&\frac{8m_{f}^{2}}{m_{B}\sqrt{4m_{f}^{2}-m_{B}^{2}}}\text{arccsc}\left(\frac{2m_{f}}{m_{B}} \right)\Bigg].
\end{eqnarray}
Setting  $\mu^2=2|eB|+m_0^2$ we obtain
\begin{eqnarray}
     \Pi_{f\bar{f}}&=&\frac{g^{2}|q_{f}B|}{2\pi^{2}}\Bigg[2\sqrt{\frac{4m_{f}^{2}}{m_B^{2}}-1}\hspace{0.1cm}\text{arccsc}\left( \frac{2m_{f}}{m_B}\right)\nonumber \\
     &+&\ln{\left(\frac{m_{f}^{2}}{2|eB|+m_0^2}\right)}\nonumber \\
     &-&\frac{8m_{f}^{2}}{m_B\sqrt{4m_{f}^{2}-m_B^{2}}}\text{arccsc}\left(\frac{2m_{f}}{m_{B}} \right)\Bigg].
\end{eqnarray}

Finally, we compute the contribution from the tadpole in Fig. \ref{self-energypions}. Its explicit expression is given by
\begin{equation}
    -i\Pi_{\pi^{\pm}}=\int \frac{d^{4}k}{(2\pi)^{4}}(-2i\lambda)iD_{\pi^{\pm}}(k),
\end{equation}
where we used the Schwinger's phase vanishes.
In the strong field limit we use the boson propagator in Eq.~(\ref{bosonpropagatorLLL}). The contribution from the two charged pions can be written as
\begin{equation}
    -i(\Pi_{\pi^{+}}+\Pi_{\pi^{-}})=-8i\lambda\int \frac{d^{4}k}{(2\pi)^{4}}\frac{i e^{-\frac{k_{\perp}^{2}}{|eB|}}}{k_{\parallel}^{2}-|eB|-m_{0}^{2}+i\epsilon}.
\end{equation}
We proceed with the integration over the perpendicular coordinates relative to the magnetic field to obtain
\begin{equation}
    -i(\Pi_{\pi^{+}}+\Pi_{\pi^{-}})= \frac{2\lambda|eB|}{\pi} \int \frac{d^{2}k_{\parallel}}{(2\pi)^{2}}\frac{1}{k_{\parallel}^{2}-\Delta_{2}+i\epsilon},
\end{equation}
where $\Delta_{2}=|eB|+m_{0}^{2}$ and $\epsilon \to 0$.
Using dimensional regularization as in Eq.~(\ref{dimensionalreg}), the integration over the parallel coordinates of momentum relative to the magnetic field direction can be found to be
\begin{eqnarray}
    \mu^{4-d}\int \frac{d^{d-2}l_{\parallel}}{(2\pi)^{d-2}}\frac{1}{l_{\parallel}^{2}-\Delta}&=&-\frac{i}{4\pi}\bigg[\frac{1}{\varepsilon}+\ln(4\pi)-\gamma_{E}\nonumber \\
    &-&\ln\left(\frac{\Delta}{\mu^{2}}\right)+\mathcal{O}(\varepsilon)\bigg].
    \label{dimensionalregularizationparallel0}
\end{eqnarray}
According to Eq.~(\ref{dimensionalregularizationparallel0}) and taking $\varepsilon\rightarrow 0$ we have
\begin{eqnarray}
    -i(\Pi_{\pi^{+}}+\Pi_{\pi^{-}})&=&\frac{-i\lambda|eB| }{2\pi^{2}}\bigg[\frac{1}{\varepsilon}+\ln(4\pi)-\gamma_{E}\nonumber\\
    &-&\ln\left(\frac{|eB|+m_{0}^{2}}{\mu^{2}}\right)\bigg].
\end{eqnarray}
Using the $\overline{MS}$ renormalization scheme, we obtain
\begin{equation}
   \Pi_{\pi^{+}}+\Pi_{\pi^{-}}=-\frac{\lambda|eB| }{2\pi^{2}}\ln\left(\frac{|eB|+m_{0}^{2}}{\mu^{2}}\right).
\end{equation}
Setting $\mu^2=2|eB|+m_0^2$, we get
\begin{equation}
   \Pi_{\pi^{+}}+\Pi_{\pi^{-}}=-\frac{\lambda|eB| }{2\pi^{2}}\ln\left(\frac{|eB|+m_{0}^{2}}{2|eB|+m_0^2}\right).
\end{equation}
\section{Magnetic corrections to the effective potential}\label{sec8}
Consider the contribution from a single charged boson to the effective potential at one-loop order, given by 
\begin{equation}
V_{b}^{1}=-\frac{i}{2}\int \frac{d^{4}k}{(2\pi)^{4}}\ln\left[-D_{b}^{-1}(k)\right],
\end{equation}
which can also be written as~\cite{villamex}
\begin{equation}
V_{b}^{1}=\frac{1}{2}\int dm_{b}^{2}\int \frac{d^{4}k}{(2\pi)^{4}}iD_{b}(k).
\end{equation}
Using the boson propagator expanded over Landau levels in Eq.~(\ref{bosonpropagatorLandauLevels}), taking a Wick rotation $k_{0}\rightarrow ik_{4}$, and using a Schwinger parameter we get
\begin{eqnarray}
V_{b}^{1}&=&\int dm_{b}^{2}\int\frac{d^{4}k_{E}}{(2\pi)^{4}}\int_{0}^{\infty}ds\sum_{n=0}^{\infty}(-1)^{n}e^{-\frac{k_{\perp}^{2}}{|q_{b}B|}} \nonumber\\
&\times& L_{n}^{0}\left( \frac{2k_{\perp}^{2}}{|q_{b}B|}\right) e^{-s[k_{E\parallel}^{2}+m_{0}^{2}+(2n+1)|q_{b}B|-i\epsilon]}.
\end{eqnarray}
Using the definitions
\begin{eqnarray}
    &&s_{1}=2k_{\perp}^{2}/|q_{b}B|, \nonumber\\ 
    &&r_{1}=-e^{-2s|q_{b}B|}, \nonumber \\
    &&\alpha(k_{E\parallel})=k_{E\parallel}^{2}+m_{b}^{2}-i\epsilon,
\end{eqnarray}
we write
\begin{eqnarray}
V_{b}^{1}&=&\int dm_{b}^{2}\int\frac{d^{4}k_{E}}{(2\pi)^{4}}\int_{0}^{\infty}ds\sum_{n=0}^{\infty}r_{1}^{n}L_{n}^{0}\left( s_{1}\right) \nonumber\\
&\times&  e^{-s[\alpha(k_{E\parallel})+|q_{b}B|]}e^{-\frac{k_{\perp}^{2}}{|q_{b}B|}}.
\end{eqnarray}
Using the generating function of Laguerre polynomials
\begin{equation}
    \sum_{n=0}^{\infty}r_{1}^{n}L_{n}^{0}(s_{1})=\frac{1}{1-r_{1}}e^{-\frac{r_{1}}{1-r_{1}}s_{1}},
\end{equation}
we have
\begin{equation}
V_{b}^{1}=\int dm_{b}^{2} \int_{0}^{\infty}ds\; \frac{e^{-s|q_{b}B|}}{1-r_{1}}I_{\perp}(s)I_{\parallel}(s),
\end{equation}
where we define 
\begin{eqnarray}
    &&I_{\perp}(s)=\int\frac{d^{2}k_{\perp}}{(2\pi)^{2}}e^{-\frac{k_{\perp}^{2}}{|q_{b}B|}(1-2\eta_{b}(s))},\nonumber\\
    &&I_{\parallel}(s)= \mu^{4-d}\int\frac{d^{d-2}k_{E\parallel}}{(2\pi)^{d-2}}e^{-s\alpha(k_{E\parallel})},\nonumber\\
    &&\eta_{b}(s)= \frac{1}{e^{2|q_{b}B|s}+1},
\end{eqnarray}
note that we can take $\epsilon \rightarrow 0$ and use dimensional regularization. Once we carry out the integration, $I_{\perp}(s)$ and $I_{\parallel}(s)$ can be written as 
\begin{eqnarray}
    &&I_{\perp}(s)=\frac{|q_{b}B|}{4\pi \tanh{(|q_{b}B|s)}},\nonumber\\
    &&I_{\parallel}(s)= \mu^{2\varepsilon}\frac{1}{(4\pi s)^{1-\varepsilon}}e^{-sm_{0}^{2}}.
    \label{IandJforbosonicpotential}
\end{eqnarray}
Using the explicit expressions in Eq.~\eqref{IandJforbosonicpotential} we get
\begin{equation}
    V_{b}^{1}=\int dm_{b}^{2} \int_{0}^{\infty}ds\;\frac{|q_{b}B|}{8\pi \sinh{(|q_{b}B|s)}}\frac{\mu^{2\varepsilon}}{(4\pi s)^{1-\varepsilon}}e^{-sm_{0}^{2}}.
\end{equation}
By writing
\begin{equation}
    \frac{1}{\sinh{(|q_{b}B|s)}}=2\sum_{n=0}^{\infty}e^{-(2n+1)|q_{b}B|s},
\end{equation}
we can perform the integration over $ds$ such that \begin{equation}
    V_{b}^{1}=\frac{|q_{b}B|}{16\pi^{2}}\int dm_{b}^{2}\;\left(\frac{4\pi\mu^{2}}{2|q_{b}B|} \right)^{\varepsilon} \Gamma(\varepsilon)\,  \zeta \left(\varepsilon,\frac{1}{2}+\frac{m_{b}^{2}}{2|q_{b}B|} \right).
\end{equation}
Considering an expansion for $\varepsilon \rightarrow 0$, we have
\begin{eqnarray}
    V_{b}^{1}&=&\frac{|q_{b}B|}{16\pi^{2}}\int dm_{b}^{2}\;\bigg\{\zeta\left(0,\frac{1}{2}+\frac{m_{b}^{2}}{2|q_{b}B|} \right)\bigg[\frac{1}{\varepsilon}-\gamma_{E}+\ln{(4\pi)}\nonumber \\
    &+&\ln{\left(\frac{\mu^{2}}{2|q_{b}B|} \right)} \bigg]+\zeta^{(1,0)}\left(0,\frac{1}{2}+\frac{m_{b}^{2}}{2|q_{b}B|} \right)\bigg\},
\end{eqnarray}
where $\zeta(s,c)$ is the Hurwitz Zeta function defined for $c>0\; \text{and}\;R[s]>1$ and by analytic continuation to other $s\neq 1$, and 
\begin{equation}
    \frac{d}{ds}\zeta(s,c)|_{s=0}=\zeta^{(1,0)}(0,c).
\end{equation}
Using the following identities 
\begin{equation}
\begin{split}
    &\zeta\left(0,c\right)=\frac{1}{2}-c, \\
    &\zeta^{(1,0)}\left(0,c\right)=\ln{\left[(2\pi)^{-1/2}\,\Gamma\left(c\right)\right]},
\end{split}
\label{identitiesZetaHurwitz}
\end{equation}
we get
\begin{eqnarray}
    V_{b}^{1}&=&-\frac{1}{32\pi^{2}}\int dm_{b}^{2}\;m_{b}^{2}\;\bigg\{\frac{1}{\varepsilon}-\gamma_{E}+\ln{(4\pi)}+\ln{\left(\frac{\mu^{2}}{2|q_{b}B|} \right)}\nonumber \\
    &-&\frac{2|q_{b}B|}{m_{b}^2}\ln{\left[(2\pi)^{-1/2}\,\Gamma\left(\frac{1}{2}+\frac{m_{b}^{2}}{2|q_{b}B|}\right)\right]}  \bigg\}.
\end{eqnarray}
Considering the $\overline{MS}$ renormalization scheme we have 
\begin{eqnarray}
    V_{b}^{1}&=&\frac{1}{32\pi^{2}}\int dm_{b}^{2}\;m_{b}^{2}\;\bigg\{\frac{2|q_{b}B|}{m_{b}^2}\ln{\left[\Gamma\left(\frac{1}{2}+\frac{m_{b}^{2}}{2|q_{b}B|}\right)\right]} \nonumber \\
    &-&\frac{|q_{b}B|}{m_{b}^2}\ln{(2\pi)}-\ln{\left(\frac{\mu^{2}}{2|q_{b}B|} \right)} \bigg\},
\end{eqnarray}
solving the integral for $dm_{b}^{2}$ we get the final result
\begin{eqnarray}
V_{b}^{1}&=&\frac{1}{16\pi^{2}}\bigg[2|q_{b}B|^{2}\psi^{-2}\left(\frac{1}{2}+\frac{m_{b}^{2}}{2|q_{b}B|}\right)-\frac{1}{2}|q_{b}B|m_{b}^{2}\ln(2\pi)\nonumber \\
&-&\frac{m_{b}^{4}}{4}\ln\left(\frac{\mu^{2}}{2|q_{b}B|} \right) \bigg],
\end{eqnarray}
where $\psi^{-2}(x)$ is the Polygamma function of order $-2$.
In the context of the LSMq we have to consider the contribution of both charged pions such as the magnetic correction from bosons in this model can be written as
\begin{eqnarray}
V_{\pi^{+}}^{1}+V_{\pi^{-}}^{1}&=&\frac{1}{8\pi^{2}}\bigg[2|eB|^{2}\psi^{-2}\left(\frac{1}{2}+\frac{m_{0}^{2}}{2|eB|}\right)\nonumber \\
&-&\frac{1}{2}|eB|m_{0}^{2}\ln(2\pi)
-\frac{m_{0}^{4}}{4}\ln\left(\frac{\mu^{2}}{2|eB|} \right) \bigg].\nonumber\\
\end{eqnarray}
The magnetic corrections from the fermion contribution to the effective potential can be computed from 
\begin{equation}
V_{f}^{1}=iN_c\int \frac{d^{4}k}{(2\pi)^{4}}{\mbox{Tr}}\; \ln\left[ S_{f}^{-1}(k)\right],
\label{Vfcont}
\end{equation}
in a similar fashion to the boson case, it can be shown that
\begin{eqnarray}
V_{f}^{1}&=&-2iN_c\sum_{\sigma=\pm 1} \int dm_{f}^{2}\int \frac{d^{4}k}{(2\pi)^{4}}\sum_{n=0}(-1)^{n}e^{-\frac{k_{\perp}^{2}}{|q_{f}B|}} \nonumber\\
&\times&\frac{L_{n}^{0}\left(\frac{2k_{\perp}^{2}}{|q_{f}B|}\right)}{k_{\parallel}^{2}-m_{f}^{2}-(2n+1+\sigma)+i\epsilon},
\end{eqnarray}
where we consider the particle-antiparticle contributions and a sum over the polarizations with
respect to the magnetic field direction, $\sigma$. Making a Wick rotation and considering a Schwinger proper time parametrization we have
\begin{eqnarray}
    V_{f}^{1}&=&-2N_c\sum_{\sigma=\pm 1}\int dm_{f}^{2}\int_{0}^{\infty} ds\int \frac{d^{4}k}{(2\pi)^{4}}\sum_{n=0}^{\infty}(-1)^{n}e^{-\frac{k_{\perp}^{2}}{|q_{f}B|}} \nonumber\\
    &\times&L_{n}^{0}\left(\frac{2k_{\perp}^{2}}{|q_{f}B|}\right)e^{-s\left[k_{E\parallel}^{2}+m_{f}^{2}+(2n+1+\sigma)|q_{f}B|-i\epsilon\right]},
\end{eqnarray}
using the following definitions 
\begin{eqnarray}
    &&s_{2}=2k_{\perp}^{2}/|q_{f}B|, \nonumber\\ 
    &&r_{2}=-e^{-2s|q_{f}B|}, \nonumber \\
    &&\beta(k_{E\parallel})=k_{E\parallel}^{2}+m_{f}^{2}-i\epsilon,
\end{eqnarray}
we get
\begin{eqnarray}
    V_{f}^{1}&=&-2N_c\sum_{\sigma=\pm 1}\int dm_{f}^{2}\int_{0}^{\infty} ds\int \frac{d^{4}k}{(2\pi)^{4}}\sum_{n=0}^{\infty}r_{2}^{n} L_{n}^{0}\left(s_{2}\right)\nonumber\\
    &\times&e^{-s\left[\beta(k_{E\parallel}^{2})+|q_{f}B|+\sigma|q_{f}B|\right]}e^{-\frac{k_{\perp}^{2}}{|q_{f}B|}}.
\end{eqnarray}
Using the generating function of Laguerre polynomials
\begin{equation}
    \sum_{n=0}^{\infty}r_{2}^{n}L_{n}^{0}(s_{2})=\frac{1}{1-r_{2}}e^{-\frac{r_{2}}{1-r_{2}}s_{2}},
\end{equation}
we obtain
\begin{eqnarray}
    V_{f}^{1}&=&-2N_c\sum_{\sigma=\pm 1}\int dm_{f}^{2}\int_{0}^{\infty} ds\int \frac{d^{4}k}{(2\pi)^{4}}\frac{e^{-s|q_{f}B|}}{1-r_{2}}\nonumber\\
    &\times&e^{-s|q_{f}B|}J_{\perp}(s)J_{\parallel}(s),
\end{eqnarray}
where after introducing dimensional regularization, we use the definitions 
\begin{eqnarray}
    &&J_{\perp}(s)=\int\frac{d^{2}k_{\perp}}{(2\pi)^{2}}e^{-\frac{k_{\perp}^{2}}{|q_{f}B|}(1-2\eta_{f}(s))},\nonumber\\
    &&J_{\parallel}(s)= \mu^{4-d}\int\frac{d^{d-2}k_{E\parallel}}{(2\pi)^{d-2}}e^{-s\beta(k_{E\parallel})},\nonumber\\
    &&\eta_{f}(s)= \frac{1}{e^{2|q_{f}B|s}+1},
\end{eqnarray}
where we have considered $\epsilon \rightarrow 0$. Once we carry out the integration, $J_{\perp}(s)$ and $J_{\parallel}(s)$ can be written as 
\begin{eqnarray}
    &&J_{\perp}(s)=\frac{|q_{f}B|}{4\pi \tanh{(|q_{f}B|s)}},\nonumber\\
    &&J_{\parallel}(s)= \mu^{2\varepsilon}\frac{1}{(4\pi s)^{1-\varepsilon}}e^{-sm_{f}^{2}}.
    \label{IandJforfermionicpotential}
\end{eqnarray}
Using the identity
\begin{equation}
    \sum_{\sigma=\pm 1}e^{-s|q_{f}B|}=2\cosh{(|q_{f}B|s)},
\end{equation}
we have
\begin{equation}
    V_{f}^{1}=-2N_c\int dm_{f}^{2}\int_{0}^{\infty} ds \frac{|q_{f}B|}{4\pi \tanh{(|q_{f}B|s)}}\frac{\mu^{2\varepsilon}}{(4\pi s)^{1-\varepsilon}}e^{-s m_{f}^{2}}.
\end{equation}
We now use that 
\begin{equation}
    \frac{1}{\tanh{(|q_{f}B|s)}}=\sum_{n=0}^{\infty}e^{-2n|q_{f}B|s}+\sum_{n=0}^{\infty}e^{-(2n+2)|q_{f}B|s},
\end{equation}
to integrate over $ds$ such that 
\begin{eqnarray}
    V_{f}^{1}&=&-\frac{N_c|q_{f}B|}{8\pi^{2}}\int dm_{f}^{2}\left(\frac{4\pi \mu^{2}}{2|q_{f}B|} \right)^{\varepsilon}\Gamma(\varepsilon)\Bigg[\zeta\left(\varepsilon,\frac{m_{f}^{2}}{2|q_{f}B|} \right) \nonumber\\ 
    &+&\zeta\left(\varepsilon,\frac{m_{f}^{2}}{2|q_{f}B|}+1 \right)\Bigg].
\end{eqnarray}
Considering an expansion when $\varepsilon \rightarrow 0$ we get
\begin{eqnarray}
    V_{f}^{1}&=&-\frac{N_c|q_{f}B|}{8\pi^{2}}\int dm_{f}^{2} \Bigg\{ \Bigg[\zeta\left(0,\frac{m_{f}^{2}}{2|q_{f}B|}+1 \right)\nonumber \\
    &+&\zeta\left(0,\frac{m_{f}^{2}}{2|q_{f}B|} \right) \Bigg] \left[\frac{1}{\varepsilon}-\gamma_{E}+\ln(4\pi)+\ln{\left(\frac{ \mu^{2}}{2|q_{f}B|} \right)} \right]\nonumber \\
    &+&\zeta^{(1,0)}\left(0,\frac{m_{f}^{2}}{2|q_{f}B|} \right)+\zeta^{(1,0)}\left(0,\frac{m_{f}^{2}}{2|q_{f}B|}+1 \right).
\end{eqnarray}
Using the identities in Eq.~\eqref{identitiesZetaHurwitz} we have
\begin{eqnarray}
    V_{f}^{1}&=&\frac{N_c}{8\pi^{2}}\int dm_{f}^{2}\; m_{f}^{2} \Bigg\{\frac{1}{\varepsilon}-\gamma_{E}+\ln(4\pi)+\ln{\left(\frac{\mu^{2}}{2|q_{f}B|} \right)} \nonumber \\
    &-&\frac{|q_{f}B|}{m_{f}^{2}}\ln{\left(\frac{m_{f}^{2}}{2|q_{f}B|}\right)}-\frac{2|q_{f}B|}{m_{f}^{2}}\ln{\left[\Gamma\left(\frac{m_{f}^{2}}{2|q_{f}B|} \right) \right]} \nonumber \\
    &+&\frac{|q_{f}B|}{m_{f}^{2}}\ln(2\pi)\Bigg\}.
\end{eqnarray}

After the $\overline{MS}$ renormalization scheme is implemented, we get
\begin{equation}
\begin{split}    
    V_{f}^{1}&=-\frac{N_c}{8\pi^{2}}\int dm_{f}^{2}\; \Bigg\{ |q_{f}B|\ln{\left(\frac{m_{f}^{2}}{2|q_{f}B|}\right)}
    -|q_{f}B|\ln(2\pi) \\
    &-m_{f}^{2}\ln{\left(\frac{\mu^{2}}{2|q_{f}B|} \right)}+2|q_{f}B|\ln{\left[\Gamma\left(\frac{m_{f}^{2}}{2|q_{f}B|} \right) \right]} \Bigg\}.
\end{split}    
\end{equation}
Finally, integrating over $dm_{f}^{2}$ we obtain
\begin{eqnarray}
V_{f}^{1}&=&-\frac{N_c}{8\pi^{2}}\bigg[4|q_{f}B|^{2}\psi^{-2}\left(\frac{m_{f}^{2}}{2|q_{f}B|}\right)-\frac{m_{f}^{4}}{2}\ln\left(\frac{\mu^{2}}{2|q_{f}B|} \right)   \nonumber \\
&-&m_{f}^{2}|q_{f}B|\left(1+\ln(2\pi)-\ln\left(\frac{m_{f}^{2}}{2|q_{f}B|} \right) \right) \bigg].
\end{eqnarray}
\vspace{6.5cm}
\end{appendix}

\end{document}